\documentclass[twocolumn,aps,prd,reprint,superscriptaddress,nofootinbib,letterpaper]{revtex4-2}

\usepackage{amsmath}
\usepackage{amsfonts}
\usepackage{amssymb}
\usepackage{graphicx, nicefrac}
\usepackage{epsfig}
\usepackage{color}
\usepackage{multirow}
\usepackage{todonotes}
\usepackage{hyperref}
\usepackage[normalem]{ulem}

\def\bea{\begin{eqnarray}}
\def\eea{\end{eqnarray}}
\def\bef{\begin{flalign}}
\def\eef{\end{flalign}}
\def\nn{\nonumber}

\def\vk{\vec{k}}
\def\vx{\vec{x}}

\def\({\left(}
\def\){\right)}
\def\[{\left[}
\def\]{\right]}
\def\<{\left\langle}
\def\>{\right\rangle}

\begin{document}


\title{Vector Perturbations in Ghost-Free Quasidilaton Massive Gravity}

\author{Ekapob Kulchoakrungsun}
\email{ek2897@nyu.edu}
\affiliation{Khon Kaen Particle Physics and Cosmology Theory Group (KKPaCT), Department of Physics, Faculty of Science, Khon Kaen University, 123 Mitraphap Rd.,
Khon Kaen, 40002, Thailand}

\author{Daris Samart}
\affiliation{Khon Kaen Particle Physics and Cosmology Theory Group (KKPaCT), Department of Physics, Faculty of Science, Khon Kaen University, 123 Mitraphap Rd.,
Khon Kaen, 40002, Thailand}

\date{\today} 

\begin{abstract}
We study transverse vector perturbations in ghost-free extended quasidilaton massive gravity without a quasidilaton kinetic term, in the presence of minimal matter. In vacuum, we recover the known result that the kinetic coefficient $K_V$ of the gravitational vector modes vanishes on the self-accelerating branch $J=0$, so those modes are infinitely strongly coupled at linear order. We then add a canonical scalar field and a single Abelian vector (Maxwell or Proca). After integrating out the auxiliary shift, we find the same $K_V$ as in vacuum. The scalar matter has no transverse perturbation; it enters the unsimplified shift constraint, but those terms cancel once the Friedmann equation is used. A Maxwell or Proca field with vanishing isotropic background does not mix with the gravitational vectors at quadratic order. Minimal matter therefore leaves $K_V=0$ on Branch~II. We do not claim that the modes are absent from the nonlinear theory. We do conclude that ordinary minimal matter is not enough to make the vector sector perturbatively healthy on this branch. If we need healthy gravitational vector modes at the linear level, Branch~I is the branch to use.

\end{abstract}

\maketitle


\section{Introduction}

The origin of the current accelerated expansion of the Universe remains a central problem in cosmology.
One possibility is an additional dark-energy component, well described by a cosmological constant; another is that gravity differs from general relativity (GR) on large scales.
Along the latter line, the graviton may be massive, with a mass of order the present Hubble scale.
Early attempts were restricted to the linear Fierz--Pauli theory~\cite{Fierz:1939ix}.
A nonlinear, ghost-free completion, the de Rham-Gabadadze-Tolley (dRGT) theory, was constructed in Refs.~\cite{deRham:2010ik,deRham:2010kj}; see Refs.~\cite{Hinterbichler:2011tt,deRham:2014zqa} for reviews.

Although free of the Boulware--Deser ghost~\cite{Boulware:1972yco}, dRGT massive gravity does not admit a viable homogeneous and isotropic Friedmann--Lema\^itre--Robertson--Walker (FLRW) cosmology~\cite{DAmico:2011eto}.
Self-accelerating FLRW solutions in dRGT suffer from vanishing kinetic terms and infinite strong coupling already at the linear level~\cite{Gumrukcuoglu:2011zh,DeFelice:2012mx}.
One proposed resolution is a quasidilaton scalar realizing an additional global symmetry~\cite{DAmico:2012hia}, with cosmological implications studied in Ref.~\cite{Gannouji:2013rwa}.
The original quasidilaton extension, however, has unstable perturbations about self-accelerating backgrounds~\cite{Gumrukcuoglu:2013nza}.
A further extension, allowing a new derivative coupling of the quasidilaton through the fiducial metric~\cite{DeFelice:2013tsa}, does admit stable self-accelerated solutions.

Background dynamics of extended quasidilaton massive gravity with a quasidilaton kinetic term were studied in Refs.~\cite{DeFelice:2013tsa,Kahniashvili:2014wua}, and cosmological perturbations in, e.g., Refs.~\cite{Motohashi:2014una,Heisenberg:2015voa,Gumrukcuoglu:2016hic}.
Away from the self-accelerating attractor the Boulware--Deser ghost is not guaranteed to remain absent~\cite{Kluson:2013jea,Mukohyama:2013raa,Anselmi:2017hwr,Golovnev:2017zxk}.
Removing the quasidilaton kinetic term yields a ghost-free subset~\cite{Gumrukcuoglu:2017ioy}.
In vacuum that theory has two FLRW branches.
On Branch~I ($\sigma'=\mathcal{H}r$) the tensor, vector, and scalar sectors can be healthy.
On Branch~II ($J=0$) the two gravitational vector modes have vanishing kinetic terms and are infinitely strongly coupled at linear order~\cite{Gumrukcuoglu:2017ioy}, analogous to the dRGT self-accelerating branch~\cite{Gumrukcuoglu:2011zh,DeFelice:2012mx}.

In Ref.~\cite{Kulchoakrungsun:2022zkk} we extended the background and \emph{scalar} perturbation analysis of this model to minimally coupled matter.
The two branches survive.
On Branch~I, demanding a positive-definite scalar kinetic matrix yields Higuchi-type bounds~\cite{Higuchi:1986py,Fasiello_2012} that are not satisfied for a generic potential, but can be satisfied when the scalar effectively behaves as pressureless matter.
In the quasistatic, subhorizon limit the growth of structure is scale-independent, as expected for a theory with a single dynamical metric.
On Branch~II the scalar equations produce a locking relation among the perturbations and remove the longitudinal St\"{u}ckelberg field from the reduced linear system.
That work did not determine whether matter changes the \emph{vector} kinetic coefficient on $J=0$.
 
The purpose of this paper is to close that gap.
We compute the quadratic vector action in vacuum, with a homogeneous canonical scalar, and with a Maxwell or Proca field whose background value vanishes.
After integrating out the auxiliary shift we obtain the same kinetic coefficient $K_V$ as in vacuum, including $K_V=0$ when $J=0$.
So minimal matter does not fix the vector sector on Branch~II at this order.
We discuss how to interpret $K_V=0$ in Sec.~\ref{section:branchII}.
If we want healthy gravitational vector modes in linear cosmology, we should work on Branch~I, where the scalar stability conditions of Ref.~\cite{Kulchoakrungsun:2022zkk} still apply.

The paper is organized as follows.
Section~\ref{section:Extended_quasidilaton_theory} reviews the ghost-free quasidilaton theory.
Section~\ref{section:Background_evolution} summarizes the FLRW background and the two branches.
Section~\ref{section:Cosmological_perturbations} introduces the transverse vector variables.
Section~\ref{section:Kinetic_matrix} computes $K_V$ in vacuum and with minimal matter.
Section~\ref{section:branchII} interprets $K_V=0$ on Branch~II.
We conclude in Sec.~\ref{section:Discussion}.

A note on our notation: Greek indices indicate time and space coordinates and take the values $0$-$3$, Latin indices indicate space coordinates and take the values $1$-$3$, and our metric signature is mostly plus. Primes denote derivatives with respect to conformal time.


\section{Extended quasidilaton theory}
\label{section:Extended_quasidilaton_theory}

The dRGT theory introduces the graviton mass in a covariant way by means of the St\"{u}ckelberg mechanism~\cite{Arkani-Hamed:2002sp,Hinterbichler:2011tt}. The four St\"{u}ckelberg fields $\phi^{\alpha}$ together generate the nondynamical metric
\bea
    f_{\mu\nu} \, = \, \eta_{\alpha\beta} \partial_{\mu} \phi^{\alpha} \partial_{\nu} \phi^{\beta} \, ,
\eea
where $\eta_{\alpha\beta}$ is the Minkowski metric and the derivative is with respect to $x^{\mu} = (t,\vx)$. The tensor $\( \sqrt{g^{-1}f} \)^{\mu}_{\ \nu}$ forms the basic building block of the mass term, $g_{\alpha\beta}$ being the dynamical spacetime metric. Since the dRGT theory does not admit a viable FLRW cosmology \cite{DAmico:2011eto}, as mentioned in the Introduction, we focus here on its quasidilaton extension, where a quasidilaton field $\sigma$ \cite{DAmico:2012hia} is coupled to the St\"{u}ckelberg fields through an extended fiducial metric \cite{DeFelice:2013tsa},
\bea
    \widetilde f_{\mu\nu}\, = \, \eta_{\alpha\beta} \partial_{\mu} \phi^{\alpha} \partial_{\nu} \phi^{\beta} - \frac{\alpha_\sigma}{m^2} \partial_{\mu} (e^{-\sigma}) \partial_{\nu} (e^{-\sigma}) \, ,
\label{eq:ftilde}
\eea
$m$ being the mass of the graviton and $\sigma$ the dimensionless field $\sigma/M_{\text{Pl}}$. This transforms as $\widetilde f_{\mu\nu} \rightarrow e^{-2\sigma_0} \widetilde f_{\mu\nu}$ under the global transformations $\sigma \rightarrow \sigma + \sigma_0$ and $\phi^{\alpha} \rightarrow e^{-\sigma_0} \phi^{\alpha}$, with $\sigma_0$ an arbitrary constant. Following \cite{Gumrukcuoglu:2017ioy}, we also remove the canonical kinetic term for the quasidilaton field.

In this paper we study the quadratic vector action of this theory in the presence of minimal matter. The action is
\bea
    S & = & \frac{M_\mathrm{Pl}^2}{2} \int d^4x \sqrt{-g} \[ R + 2m^2 \( \mathcal{L}_2 + \alpha_3 \mathcal{L}_3 + \alpha_4 \mathcal{L}_4 \) \] \nn \\
    & & \quad + \, \int d^4 x \sqrt{-g} \, \mathcal{L}_\mathrm{matter} \,,
\label{eq:action}
\eea
where $R$ is the 4D Ricci scalar. The mass term is generated by the Lagrangian densities
\bea
    \mathcal{L}_2 & = & \frac{1}{2!} \( [\mathcal{K}]^2 - [\mathcal{K}^2] \) , \\
    \mathcal{L}_3 & = & \frac{1}{3!} \( [\mathcal{K}]^3 - 3[\mathcal{K}][\mathcal{K}^2] + 2 [\mathcal{K}^3] \) , \\
    \mathcal{L}_4 & = & \frac{1}{4!} \big( [\mathcal{K}]^4 - 6[\mathcal{K}]^2[\mathcal{K}^2] + 3 [\mathcal{K}^2]^2 + 8[\mathcal{K}][\mathcal{K}^3] \nn \\
    & & \quad - \ 6[\mathcal{K}^4] \big) \, ,
\eea
where square brackets denote the trace, and
\bea
    \mathcal{K}^\mu_{\ \nu} \, = \, \delta^\mu_\nu - e^{\sigma} \( \sqrt{g^{-1} \widetilde f} \)^{\mu}_{\ \nu} \, .
\eea
The matter Lagrangian density $\mathcal{L}_\mathrm{matter}$, on the other hand, gives the energy-momentum tensor in the standard way,
\bea
    T_{\mu\nu} & = & -\frac{2}{\sqrt{-g}} \frac{\delta}{\delta g^{\mu\nu}} \( \sqrt{-g} \, \mathcal{L}_\mathrm{matter} \) .
\label{eq:stress-energy tensor}
\eea
We use the action in Eq.~(\ref{eq:action}) for the background, which we review in the next section, and for the vector perturbations of Secs.~\ref{section:Cosmological_perturbations} and~\ref{section:Kinetic_matrix}.
Matter is minimally coupled to $g_{\mu\nu}$ only.


\section{Background evolution}
\label{section:Background_evolution}

Let us first consider the Einstein-Hilbert part of the action. For this, we choose to work in the Arnowitt-Deser-Misner (ADM) formalism, since this allows us to easily identify the boundary term, as explained further below. In this formalism, spacetime is foliated into spacelike hypersurfaces and is described in terms of a spatial metric $h_{ij}$, lapse function $N$, and shift vector $N^i$, with the spacetime interval written as
\bea
    ds^2 & = & -N^2 dt^2 + h_{ij} \(N^i dt + dx^i \) \( N^j dt + dx^j \) .
\eea
The functions here are related to components of the original metric $g_{\mu\nu}$ as
\bea
    g_{\mu\nu} \, = \, \( {\begin{array}{cc}
        -N^2 + N^k N_k & N_j \\
        N_i & h_{ij} \\
    \end{array}} \) ,
\eea
while for the inverse metric, we have
\bea
    g^{\mu\nu} \, = \, \frac{1}{N^2} \( {\begin{array}{cc}
        1 & N^j \\
        N^i & N^2 h^{ij} - N^i N^j \\
    \end{array}} \) .
\eea
We can now use the Gauss-Codazzi equation to relate the 4D Ricci scalar to the 3D Ricci scalar $R^{(3)}$,
\bea
    R \, = \, R^{(3)} + K^i_j K^j_i - [K]^2 + 2\nabla_{\mu} \( [K] n^{\mu} - n^{\nu} \nabla_{\nu} n^{\mu} \) , \nn \\
\label{eq:gc}
\eea
where $\nabla_{\mu}$ is the covariant derivative with respect to the metric $g_{\mu\nu}$, $K_{ij} = \frac{1}{2N} \big( \dot{h}_{ij} - D_i N_j - D_j N_i \big)$ is the extrinsic curvature, with the dot denoting a derivative with respect to $t$ and $D_i$ being the covariant derivative with respect to the induced metric $h_{ij}$, and $n^{\mu} = \frac{1}{N} \( 1, -N^i \)$ is the normal to the spatial slice. The boundary term is the last term in Eq.~(\ref{eq:gc}) and is cancelled by the Gibbons-Hawking-York boundary term~\cite{York:1972sj,Gibbons:1977uc}.
Identifying it this way makes the second-order action unambiguous.
For the background we take flat FLRW, $N_i=0$ and $h_{ij}=a^2(t)\delta_{ij}$, where $a(t)$ is the scale factor.

Let us next consider the massive gravity part of the action. Consistent with a homogeneous and isotropic background cosmology, we choose the background St\"{u}ckelberg fields to be 
\bea
    \phi^{\alpha} & = & \delta^{\alpha}_0 \phi(t) + \delta^{\alpha}_i x^i \, ,
\eea
where $\phi(t)$ is some function of $t$. Note that we will not work in unitary gauge for the St\"{u}ckelberg fields and will thus allow their perturbations to be functions of spacetime later in the paper. The background quasidilaton field is similarly chosen to be a function of $t$ only. Using this in Eq.\ (\ref{eq:ftilde}) gives the following fiducial spacetime interval,
\bea
    ds^2_{\tilde{f}} & = & -r^2(N^2/a^2) dt^2 + \delta_{ij} dx^i dx^j \, ,
\eea
where
\bea
    r^2 \frac{N^2}{a^2} & = & \dot{\phi}^2 + \frac{\alpha_\sigma}{m^2} e^{-2\sigma} \dot{\sigma}^2
\label{eq:r definition}
\eea
is the effective lapse function.

We can now obtain the zeroth order action from Eq.\ (\ref{eq:action}). We define $X = e^{\sigma}/a$ and the following combinations of background quantities as in \cite{Gumrukcuoglu:2017ioy},
\bea
    J & = & (3-2X) + (X-3)(X-1) \alpha_3 + (X-1)^2 \alpha_4 \, , \nn \\
\label{eq:def J} \\
    Q & = & (X-1) [3 - 3(X-1) \alpha_3 + (X-1)^2 \alpha_4] \, ,
\label{eq:def Q}
\eea
in terms of which the zeroth order action is given by
\bea
    S^{(0)} & = & M_{\rm Pl}^2 \int dt d^3x a^3 N \[ -3H^2 + m^2 (rQX - \rho_X) \] \nn \\ 
    & & \quad + \ S_{\rm matter} \, ,
\eea
where $H = \dot{a}/(aN)$ is the Hubble parameter and
\bea
    \rho_X & = & \frac{1}{X} \[ Q + J (X-1)^2 - X (X-1)^2 \]
\eea
can be interpreted as the contribution to the energy density from massive gravitons. The background equations of motion are obtained by varying $S^{(0)}$ with respect to the background fields $\{N(t), a(t), \phi(t), \sigma(t)\}$,
\bea
    \delta S^{(0)} & = & \int d^4x \sum_i \( \frac{\delta S^{(0)}}{\delta\Phi^i} \delta\Phi^i \) \, ,
\label{eq:varying action}
\eea
and setting the variation $\delta S^{(0)}$ to zero. (The summation here is over the four fields mentioned above and not a spatial index.) After the variation, we can either set $N$ to unity, so that the time coordinate corresponds to physical time $t$, or to $a(\tau)$, so that the time coordinate corresponds to conformal time $\tau$. We will work in conformal time below.

Assuming that matter is a perfect fluid so that its energy-momentum tensor is given by $\bar{T}^{\mu}_{\nu} = {\rm diag} ( -\bar{\rho}, \bar{p}, \bar{p}, \bar{p} )$, where a bar denotes background quantities, the resulting four independent equations describing the background evolution are 
%
\bea
     \frac{\mathcal{H}^2}{a^2} & = & \frac{m^2 \rho _X}{3} + \frac{\Bar{\rho}}{ 3M_{\text{Pl}}^2} \, ,
\label{eq:friedmann1}
\eea
%
\bea
    2\( \frac{\mathcal{H}' - \mathcal{H}^2}{a^2} \) & = & m^2 JX(r-1) - \frac{\bar{\rho} + \bar{p}}{M_{\mathrm{Pl}}^2} \, ,
\label{eq:friedmann2}
\eea
%
\bea
    \frac{d}{d\tau} \( \frac{a^4 QX\phi'}{r} \) & = & 0 \, ,
\label{eq:Stuckelberg background}
\eea
%
\bea
    \frac{\alpha_\sigma}{X a^5} \frac{d}{d\tau} \( \frac{a^3 Q \sigma'}{r} \) & = & m^2 X [3J(r-1) + 4rQ] \, , \quad
\label{eq: sigma background}
\eea
where ${\cal H} = aH$ is the conformal Hubble parameter and a prime denotes a derivative with respect to $\tau$. Combining the first and second Friedmann equations\ (\ref{eq:friedmann1}) and (\ref{eq:friedmann2}) and using the relationships $\bar{\rho}' + 3{\cal H}(\bar{\rho} + \bar{p}) = 0$, $\rho_X' = 3JX'$, and $X' = (\sigma' - {\cal H}) X$ yields the constraint
\bea
    m^2 J X \left( \sigma' - \mathcal{H}r \right) & = & 0 \, .
\label{eq:two branches eq}
\eea
This restricts the background to the same two branches found in vacuum~\cite{Gumrukcuoglu:2017ioy}:
\bea
    \textrm{Branch~I:}
    &&
    \sigma' \, = \, \mathcal{H} r \,,
    \quad
    X' \, = \, (r-1)\mathcal{H} X \,,
    \\
    \textrm{Branch~II:}
    &&
    J \, = \, 0 \,,
    \quad
    X' \, = \, 0 \,,
    \quad
    \sigma' \, = \, \mathcal{H} \, .
\eea
Matter does not generate a third branch.
On Branch~I the scalar sector with minimal matter was analyzed in Ref.~\cite{Kulchoakrungsun:2022zkk}: the kinetic matrix can be positive definite for a restricted class of potentials, and subhorizon growth is scale-independent.
On Branch~II, $X$ is locked to a constant root of $J(X)=0$.
The scalar analysis of Ref.~\cite{Kulchoakrungsun:2022zkk} then yields a linear locking relation, while the vector kinetic term vanishes in vacuum~\cite{Gumrukcuoglu:2017ioy}.
Whether that vanishing survives minimal matter is the subject of the rest of this paper.


\section{Transverse Vector Perturbations}
\label{section:Cosmological_perturbations}

We now turn to linear vector perturbations about the FLRW backgrounds of Sec.~\ref{section:Background_evolution}.
In this section we introduce the perturbation variables, explain why the scalar, vector, and tensor sectors decouple, fix the residual vector gauge freedom, and identify which fields are dynamical.
The quadratic action and the resulting kinetic coefficients $K_V$ are derived in Sec.~\ref{section:Kinetic_matrix}, first in vacuum and then with minimal matter.

\subsection{Metric, St\"{u}ckelberg, and matter variables}
 
A general linearized perturbation of a spatially flat FLRW metric can be decomposed, using rotational invariance of the background, into scalar, transverse-vector, and transverse-traceless tensor pieces (the standard SVT decomposition~\cite{Bardeen:1980kt,Kodama:1984ziu,Ma:1995ey}).
Explicitly,
\bea
    ds^2 &=& a^2(\tau) \bigg[ -(1+2\Phi)\,d\tau^2 + 2\bigl( \partial_i B + B_i \bigr)\,dx^i d\tau \nn \\
    && + \bigg\{ (1-2\Psi)\delta_{ij} + 2\Bigl( \partial_i\partial_j - \tfrac{\delta_{ij}}{3}\partial^2 \Bigr) E \nn \\
    &&
    \qquad + \partial_{(i} E_{j)} \bigg\} dx^i dx^j \bigg] \,,
\label{Eq:metric_perturbation}
\eea
where $\Phi$, $B$, $\Psi$, and $E$ are scalar metric perturbations, $B_{i}$ and $E_{i}$ are transverse vector perturbations, and we have omitted the tensor modes $\gamma_{ij}$, since they are not needed for the analysis in this paper. All perturbations are functions of $(\tau,\vx)$.
All three-vectors are defined to be divergenceless,
\bea
    \partial^{i}B_{i} & = & 0\, , \\
    \partial^{i}E_{i} & = & 0\,.
\label{eq:vector_divfree}
\eea
These conditions remove the scalar parts of $B_i$ and $E_i$ that would otherwise be redundant with $B$ and $E$.
In components, $B_i$ is the vector part of the shift and $E_i$ parametrizes a vector distortion of the spatial metric.
We do not study tensors in this paper: they decouple from the vector sector at linear order, and including them would not affect the kinetic coefficient of the vector modes derived below.
 
Because the graviton mass term is built from $g^{-1}\widetilde f$, we must also perturb the St\"{u}ckelberg fields that define $\widetilde f_{\mu\nu}$.
We do not impose unitary gauge on $\phi^\alpha$, so their perturbations are retained as explicit fields.
Writing
\bea
    \delta\phi^0 & = & \Pi^0 \, , \\
    \delta\phi^i & = & \Pi^i + \partial^i \Pi_L \, ,
\label{eq:stuckelberg_pert}
\eea
we split $\delta\phi^i$ into a transverse vector $\Pi^i$ and a longitudinal scalar $\Pi_L$, with
\bea
    \partial_{i}\Pi^{i} \, = \, 0\,.
\eea
The background choice $\phi^i=x^i$ makes $\Pi^i$ the helicity-1 Goldstone mode associated with broken spatial diffeomorphisms.
The quasidilaton perturbation $\delta\sigma(\tau,\vx)$ is a scalar.
 
Matter, when present, is decomposed in the same way.
A canonical scalar $\chi=\bar\chi(\tau)+\delta\chi$ contributes only a scalar perturbation $\delta\chi$; it has no transverse vector mode about a homogeneous background.
A single Abelian vector field $A_\mu$ with vanishing isotropic background, $\bar A_\mu=0$, contributes a transverse spatial perturbation $A_i^T$ in the vector sector, together with scalar pieces $(A_0,A_L)$ that we do not need here.
We write
\bea
\delta A_i \, = \, A_i^T + \partial_i A_L \,,
\qquad
\partial^i A_i^T \, = \, 0 \, .
\eea
If the vector is massive, the split is the same. The Proca term can give $A_i^T$ a mass, but it does not add a new gravitational vector field.
 
\subsection{Decoupling on FLRW}
 
The background is homogeneous and isotropic.
At linear order, the quadratic action is therefore invariant under spatial rotations, and fields transforming as different SO$(3)$ representations cannot mix.
In particular, a scalar (helicity $0$), a divergenceless vector (helicity $\pm 1$), and a transverse-traceless tensor (helicity $\pm 2$) appear in separate quadratic Lagrangians.
This is the standard SVT decoupling of cosmological perturbation theory; it continues to hold in the present theory because the FLRW background, including the homogeneous St\"{u}ckelberg and quasidilaton configurations of Sec.~\ref{section:Background_evolution}, preserves the same symmetries.
 
Consequently, when we construct the second-order vector action we may set every scalar and tensor perturbation to zero without loss of generality,
\bea
    \Phi=B=\Psi=E=\Pi^0=\Pi_L=\delta\sigma=\delta\chi=0\,,
\label{eq:set_scalars_to_zero}
\eea
and likewise $\gamma_{ij}=0$.
The remaining gravitational variables are $(B_i,E_i,\Pi^i)$.
If an Abelian matter field is included, $A_i^T$ is retained as well.
 
This decoupling holds at linear order on FLRW.
Cubic and higher vertices can mix sectors; for example a scalar can couple to two vector modes. We come back to that only when we discuss $J=0$ in Sec.~\ref{section:branchII}.

\subsection{Vector gauge freedom}

Even after applying the SVT decomposition, a residual gauge freedom remains in the vector sector.
An infinitesimal coordinate transformation $x^\alpha\rightarrow x^\alpha+\xi^\alpha$ with a transverse spatial parameter $\xi_i$ ($\partial^i\xi_i=0$) does not generate scalar or tensor perturbations.
The metric and St\"{u}ckelberg vectors transform as
\bea
    B_i &\rightarrow& B_i - \xi_i' \,, \label{eq:vector_gauge1} \\
    E_i &\rightarrow& E_i - 2\xi_i \,, \label{eq:vector_gauge2} \\
    \Pi_i &\rightarrow& \Pi_i - \xi_i \,. \label{eq:vector_gauge3}
\eea
The factor of $2$ in the rule for $E_i$ follows from our normalization in Eq.~(\ref{Eq:metric_perturbation}): the spatial metric contains $\partial_{(i}E_{j)}$, while a spatial diffeomorphism shifts that component by $-\partial_i\xi_j-\partial_j\xi_i=-2\partial_{(i}\xi_{j)}$.
The rule for $\Pi_i$ follows from the background $\phi^i=x^i$, so that a spatial reparametrization is equivalent to a shift of $\delta\phi^i$.
 
These transformations make it clear that $B_i$, $E_i$, and $\Pi_i$ are not all independent physical fields.
The combinations
\bea
    B_i^{\mathrm{GI}} \, = \, B_i - \frac12 E_i' \, ,
    \qquad
    \Pi_i^{\mathrm{GI}} \, = \, \Pi_i - \frac12 E_i
\label{eq:vector_GI}
\eea
are invariant under Eqs.~(\ref{eq:vector_gauge1})-(\ref{eq:vector_gauge3}).
We can therefore either work exclusively with $B_i^{\mathrm{GI}}$ and $\Pi_i^{\mathrm{GI}}$, or fix a gauge that eliminates one of the original fields.
 
We adopt the latter option for simplicity and choose the spatial vector gauge
\bea
    E_i \, = \, 0 \, .
\label{eq:vector_gauge_choice}
\eea
This is always available: starting from a generic $E_i$, the choice $\xi_i=E_i/2$ sets $\widetilde E_i=0$, after which no residual transverse diffeomorphism remains.
In this gauge,
\bea
    B_i^{\mathrm{GI}} \, = \, B_i \,,
    \qquad
    \Pi_i^{\mathrm{GI}} \, = \, \Pi_i \,,
\eea
so the fields that appear in the action below are already gauge invariant.
If a Maxwell field is present, $A_i^T$ is itself gauge invariant on the $\bar A_\mu=0$ background (up to the usual $U(1)$ redundancy, which does not mix with $\xi_i$).
Any additional matter-sector gauge choice, such as unitary gauge for a scalar field, will be stated when that sector is introduced; such a choice lives entirely in the scalar sector and does not affect the vector analysis.
 
\subsection{Fourier space and auxiliary fields}

We expand all perturbations in Fourier modes,
\bea
    f(\tau,\vx) \, = \, \int \frac{d^3k}{(2\pi)^3}\, e^{i \vk \cdot \vx}\, f(\tau,\vk) \,,
\label{eq:fourier}
\eea
and write $k=|\vk|$.
The divergenceless conditions then become $\vk\cdot\mathbf{B}=\vk\cdot\mathbf{E}=\vk\cdot\boldsymbol{\Pi}=0$, so each vector carries two independent polarizations for $k\neq 0$.
Because the background is isotropic, the two helicities share the same quadratic action; it is enough to keep track of a single copy and restore the sum over polarizations at the end if desired.
 
As in the ADM treatment of GR, the lapse and shift appear in the Einstein--Hilbert term without time derivatives.
The same is true of the vector shift $B_i$ once the full second-order action, the Einstein--Hilbert plus the dRGT-type mass term, is assembled. $B_i'$ has no kinetic term and is therefore auxiliary.
The equation of motion for $B_i$ is an algebraic constraint, which can be solved for $B_i$ in terms of $\Pi_i'$ (and, if they enter, matter vectors).
Substituting that solution back into the action yields the reduced quadratic action for the propagating helicity-1 fields.
 
In the next section we always proceed in the same way.
We set $E_i=0$ and use Eq.~(\ref{eq:set_scalars_to_zero}), keep only the vector part of $S^{(2)}$, solve the $B_i$ constraint, and read off the kinetic coefficient of $\Pi_i$, which we call $K_V$.
We first repeat the vacuum calculation of Ref.~\cite{Gumrukcuoglu:2017ioy} and then add a canonical scalar and a Maxwell or Proca field with $\bar A_\mu=0$.
Long intermediate expressions that are not needed for $K_V$ are left in the {\tt Mathematica} supplement.


\section{Quadratic action and kinetic coefficients}
\label{section:Kinetic_matrix}

We now follow the four steps listed at the end of Sec.~\ref{section:Cosmological_perturbations}.
After the SVT decomposition and the gauge fixing $E_i=0$, the gravitational vector variables are the auxiliary shift $B_i$ and the St\"{u}ckelberg vector $\Pi_i$.
We expand Eq.~(\ref{eq:action}) to second order, isolate the vector sector, solve the $B_i$ constraint, and read off the kinetic coefficient $K_V$ of $\Pi_i$.

For the gravitational vector modes to propagate healthily at this order, $K_V$ should be finite and have a definite sign.
If it vanishes on an entire cosmological branch, those modes have no quadratic kinetic term and are infinitely strongly coupled~\cite{Gumrukcuoglu:2017ioy}.
We first recover the vacuum result of Ref.~\cite{Gumrukcuoglu:2017ioy} and then add a scalar field and an Abelian vector.
We find the same $K_V$ throughout, including $K_V=0$ on Branch~II.
A discussion of the infinite strong coupling is given in Sec.~\ref{section:branchII}.

We do not write the mass/gradient piece of the reduced vector action.
It is not needed to diagnose whether $K_V$ vanishes. The unsimplified $B_i$ constraint with scalar matter is displayed in Sec.~\ref{sec:vectors_scalar_matter}, because that is where matter first appears.
\subsection{Vacuum benchmark}
\label{sec:vectors_vacuum}

For completion and as a check of the setup, we first repeat the vacuum calculation of Ref.~\cite{Gumrukcuoglu:2017ioy}.
We use the same gauge, $E_i = E = \delta \sigma = 0$, which is already contained in Eqs.~(\ref{eq:set_scalars_to_zero}) and~(\ref{eq:vector_gauge_choice}).
The only gravitational vector variables are then $B_i$ and $\Pi_i$.

The quadratic action contains no time derivative of $B_i$, so $B_i$ is auxiliary and its equation of motion is algebraic.
Solving that constraint before any background equation is used gives
\begin{widetext} 
\bea
    B_i = \frac{2 a^2 J m^2 X^2 \, \Pi _i' }{ (1+r) X ( k^2 + 6 \mathcal{H}^2)  -2 a^2 m^2 \{ J \left[r (X-1)^2-2 X+1\right]+(r+1) \left[Q-(X-1)^2 X\right]\} }  \, .
    \label{eq:B_i_with_Q}
\eea
\end{widetext}
The combination $Q$ is not independent of the other background quantities.
In vacuum, the Friedmann equation together with the definition of $\rho_X$ implies
\bea
    Q
    =
    \frac{3 X \mathcal{H}^2}{a^2 m^2}
    -(X-1)^2(J-X) \, .
\label{eq:Q_vacuum}
\eea
Substituting Eq.~(\ref{eq:Q_vacuum}) into Eq.~(\ref{eq:B_i_with_Q}) simplifies the constraint to
\bea
    B_i
    =
    \frac{2 a^2 J m^2 X \,\Pi_i'}
    {(1+r)k^2 + 2 a^2 J m^2 X} \, ,
\label{eq:B_i_vacuum}
\eea
in agreement with Ref.~\cite{Gumrukcuoglu:2017ioy}.
The right-hand side is proportional to $J$.
On Branch~II we have $B_i=0$.

Substituting Eq.~(\ref{eq:B_i_vacuum}) back into the quadratic action produces a reduced action for $\Pi_i$ alone.
Its kinetic term can be written as
\bea
    S_{V,\mathrm{kin}}^{(2)}
    =
    \int d\tau\,\frac{d^3k}{(2\pi)^3}\,
    \frac{a^2 M_{\mathrm{Pl}}^2 k^2}{4}\,
    K_V\,
    \Pi_i^{\prime *}\Pi_i' \,,
\label{eq:S_V_kin}
\eea
where
\bea
    K_V
    =
    \left[
    1+\frac{k^2(r+1)}{2 a^2 J m^2 X}
    \right]^{-1}
    =
    \frac{2 a^2 J m^2 X}
    {2 a^2 J m^2 X + k^2(r+1)} \, .
\label{eq:K_V}
\eea
This is the vacuum kinetic coefficient obtained in Ref.~\cite{Gumrukcuoglu:2017ioy}.

On Branch~I, $J$ need not vanish.
Then, for $X>0$ and $r+1>0$, $K_V$ is nonzero, and the ultraviolet no-ghost condition reduces to $J>0$:
\bea
    a^2 k^2 K_V
    \;\xrightarrow{k\rightarrow\infty}\;
    \frac{2 a^4 m^2 J X}{r+1} \, .
\eea
We do not study Branch~I further in this paper.
On Branch~II,
\bea
    J=0
    \qquad\Longrightarrow\qquad
    K_V=0 \, .
\label{eq:KV_J0}
\eea
Both gravitational vector polarizations then have no quadratic kinetic term, and are infinitely strongly coupled as identified in Ref.~\cite{Gumrukcuoglu:2017ioy}.
The remainder of this section asks whether adding minimal matter can restore a nonzero $K_V$ on the same branch.

\subsection{Canonical scalar matter}
\label{sec:vectors_scalar_matter}

The simplest matter sector that sources a nontrivial FLRW background is a canonical scalar field,
\bea
    \mathcal{L}_{\mathrm{matter}}
    & = &
    -\frac12 \bigl(\partial_{\mu}\chi\bigr)^2 - V(\chi) \,,
\label{eq:scalar_matter}
\eea
where $(\partial_{\mu} \chi)^2 = g^{\mu\nu} (\partial_{\mu}\chi) (\partial_{\nu}\chi)$ and $V$ is some potential. We perturb the scalar field as $\chi=\bar\chi(\tau)+\delta\chi(\tau,\vx)$, where $\bar{\chi}(\tau)$ is the background field.
The background energy density and pressure are
\bea
    \bar\rho
    & = &
    \frac{(\bar\chi')^2}{2a^2}+V(\bar\chi) \,,
    \qquad
    \bar p
    =
    \frac{(\bar\chi')^2}{2a^2}-V(\bar\chi) \,,
\label{eq:scalar_rho_p}
\eea
so that the background equations remain Eqs.~(\ref{eq:friedmann1}) and~(\ref{eq:friedmann2}), now with this $\bar\rho$ and $\bar p$.

Because $\delta\chi$ is a scalar, it has no transverse component about a homogeneous background and does not appear in the linear vector action.
There is therefore no additional propagating vector from the matter field, and no scalar-sector gauge choice is required for this calculation.
The only way $\chi$ can affect $K_V$ is indirectly, through background quantities that enter the coefficients of the gravitational vector action.

The constraint for $B_i$ remains algebraic.
Before the background equations are used, its solution is
\begin{widetext}
\bea
    B_i
    & = &
    \frac{2 a^2 J m^2 X^2 M_{\mathrm{Pl}}^2 }
    {\mathcal{D}_{\chi}}\,\Pi_i' \,,
\label{eq:B_i_scalar_raw}
\eea
with
\bea
    \mathcal{D}_{\chi}
    & = &
    -2 a^2
    \Big\{
    J m^2 M_{\mathrm{Pl}}^2 \bigl[r(X-1)^2-2X+1\bigr]
    +(r+1)
    \bigl[
    m^2 M_{\mathrm{Pl}}^2 \bigl(Q-(X-1)^2 X\bigr)
    + X V(\bar\chi)
    \bigr]
    \Big\}
    \nn \\
    &&
    +(r+1) X \bigl( k^2 M_{\mathrm{Pl}}^2 - (\bar\chi')^2 \bigr)
    + 6(r+1) X \mathcal{H}^2 M_{\mathrm{Pl}}^2 \, .
\label{eq:D_chi}
\eea
\end{widetext}
Compared with Eq.~(\ref{eq:B_i_with_Q}), the new ingredients are $V(\bar\chi)$ and $\bar\chi'$.
They are not independent of the gravitational background: the Friedmann equation~(\ref{eq:friedmann1}) relates $V(\bar\chi)$, $(\bar\chi')^2$, and $\mathcal{H}^2$ to $\rho_X$.
After that substitution, every explicit matter term in $\mathcal{D}_{\chi}$ cancels and Eq.~(\ref{eq:B_i_scalar_raw}) reduces exactly to the vacuum constraint, Eq.~(\ref{eq:B_i_vacuum}).
Substituting this $B_i$ back into the action then yields the same kinetic term as in vacuum, Eqs.~(\ref{eq:S_V_kin}) and~(\ref{eq:K_V}).

A homogeneous canonical scalar can change the background evolution of $a$, $r$, and $X$.
It does not change the functional dependence of $K_V$ on those quantities.
In particular,
\bea
    J \, = \, 0
    \qquad\Longrightarrow\qquad
    K_V \, = \, 0
\label{eq:KV_J0_scalar}
\eea
still holds.
Minimal scalar matter does not restore a quadratic kinetic term for $\Pi_i$ on Branch~II, so the infinite strong-coupling problem remains.

\subsection{Abelian vector matter}
\label{sec:vectors_vector_matter}

A canonical scalar is not a decisive test of the vector kinetic term, because $\delta\chi$ never enters that sector.
A more interesting question is whether a matter field that *does* carry a transverse vector mode can mix with $\Pi_i$ and regenerate a nonzero $K_V$ on Branch~II.
To address this, we couple the theory minimally to a single Abelian vector $A_{\mu}$, first massless (Maxwell) and then massive (Proca).
In both cases, the matter field interacts with gravity only through $g_{\mu\nu}$; it does not couple directly to the St\"{u}ckelberg fields or to the quasidilaton.

\paragraph{Background.}
A single Abelian field cannot support a nonzero homogeneous and isotropic field strength on FLRW.
We therefore take
\bea
    \bar A_{\mu} \, = \, 0 \,,
\label{eq:Abar_zero}
\eea
which for a Maxwell field can also be viewed as a residual $U(1)$ choice.
With this background the vector field does not contribute to $\bar\rho$ or $\bar p$, so it does not modify the FLRW equations of Sec.~\ref{section:Background_evolution}.
The vector is then only a perturbation: any additional perfect-fluid or scalar content may still source the background, but the vector itself does not.

Linear perturbations are then $A_{\mu}$ itself.
As in Eq.~(\ref{eq:stuckelberg_pert}), we split the spatial part into transverse and longitudinal pieces,
\bea
    \delta A_i \, = \, A_i^T + \partial_i A_L \,,
    \qquad
    \partial^i A_i^T \, = \, 0 \, .
\eea
Only $A_i^T$ belongs to the vector sector.
The components $A_0$ and $A_L$ are scalars. When $m_A\neq 0$, the extra Proca degree of freedom sits in that scalar sector, so it cannot give $\Pi_i$ a kinetic term.

\paragraph{Maxwell field.}
The Maxwell action is
\bea
    S_{A}
    & = &
    -\frac14 \int d^4x \sqrt{-g}\, F_{\mu\nu} F^{\mu\nu} \,, \\
    F_{\mu\nu}
    & = &
    \partial_{\mu}A_{\nu}-\partial_{\nu}A_{\mu} \, .
\label{eq:maxwell_action}
\eea
Because $\bar A_{\mu}=0$, the field strength starts at first order in perturbations.
The quadratic part of Eq.~(\ref{eq:maxwell_action}) is therefore built from two powers of $\delta A$ and the background metric only.
A vertex that also involves a metric perturbation $\delta g_{\mu\nu}$ (hence $B_i$ or $E_i$) contains one power of $\delta g$ and two powers of $\delta A$, so it is cubic.
There is likewise no quadratic mixing with $\Pi_i$, because $A_{\mu}$ couples only to $g_{\mu\nu}$.

In conformal time the transverse quadratic action reduces to the standard flat-space form, a consequence of the conformal invariance of Maxwell theory in four dimensions,
\bea
    S_{A,V}^{(2)}
    & = &
    \frac12 \int d\tau\,\frac{d^3k}{(2\pi)^3}
    \left[
    A_i^{T\prime *} A_i^{T\prime}
    - k^2 A_i^{T*} A_i^{T}
    \right] .
\label{eq:maxwell_vector_action}
\eea
The gravitational vector calculation is then identical to the vacuum one.
The $B_i$ constraint is still Eq.~(\ref{eq:B_i_with_Q}) and, after the background equations are used, Eq.~(\ref{eq:B_i_vacuum}).
The complete quadratic vector action is the direct sum of the gravitational piece, whose kinetic term is Eq.~(\ref{eq:S_V_kin}), and Eq.~(\ref{eq:maxwell_vector_action}).
In the basis $(\Pi_i,A_i^T)$ the kinetic matrix is therefore diagonal:
one eigenvalue is proportional to $K_V$, and the other is the healthy Maxwell kinetic term.
A vanishing $K_V$ is not compensated by the photon; the two directions are orthogonal at quadratic order.

\paragraph{Proca field.}
Adding a Proca mass,
\bea
    S_{\mathrm{Proca}}
    & = &
    S_{A}
    - \frac12 m_A^2 \int d^4x \sqrt{-g}\, A_{\mu} A^{\mu} \,,
\label{eq:proca_action}
\eea
does not change this counting when $\bar A_{\mu}=0$.
The mass term is quadratic in $A_{\mu}$.
On the vanishing background its quadratic expansion uses only the FLRW metric and produces a mass for $A_i^T$ (and for $A_0$, $A_L$).
A mixing of the form $B_i A_i^T$ or $\Pi_i A_i^T$ would require one power of $\bar A_{\mu}$ and is absent.
The $B_i$ constraint and the gravitational kinetic coefficient are therefore the same as in vacuum, Eq.~(\ref{eq:K_V}).
The only new quadratic ingredient is a mass term for $A_i^T$ in Eq.~(\ref{eq:maxwell_vector_action}).
The limit $m_A\rightarrow 0$ recovers Maxwell, as a check.

In short, the Proca term only gives a mass to the matter vector.
It does not restore a kinetic term for $\Pi_i$.

\paragraph{Branch~II.}
On Branch~II we still have $K_V=0$.
The photon, or the massive Proca vector, can propagate, but it is a matter mode.
The gravitational kinetic eigenvalue is still proportional to $K_V$ and therefore vanishes, so the gravitational vector modes remain without a quadratic kinetic term.

This conclusion is tied to $\bar A_{\mu}=0$.
A nonzero isotropic vector background can produce quadratic mixing, but it cannot be realized with a single Abelian field. We would need several vectors, or a non-Abelian configuration, which we do not consider here.


\section{Interpretation of the \texorpdfstring{$J=0$}{J=0} branch}
\label{section:branchII}

In Sec.~\ref{section:Kinetic_matrix} we found, with and without minimal matter, that
\bea
    J \, = \, 0
    \qquad\Longrightarrow\qquad
    K_V \, = \, 0 \, .
\eea
This section explains what that fact does and does not imply.
The distinction matters, because the same algebra can be read in two ways: as a linear signal of infinite strong coupling, or as evidence that the vector modes have been removed from the theory.
Only the first reading is justified by the present calculation.

On Branch~II the background condition $J=0$ is not reached asymptotically; it is imposed at every time.
For generic parameters it fixes $X$ to a constant root $X_\star$ of Eq.~(\ref{eq:def J}), and therefore locks the homogeneous quasidilaton to the scale factor,
\bea
    X \, = \, X_\star \,,
    \qquad
    \sigma' \, = \, \mathcal{H} \, .
\label{eq:branchII_lock}
\eea
Thus $K_V=0$ holds for the whole background, not just at one scale or one time.

When $K_V=0$, the quadratic action for $\Pi_i$ has no time-derivative term.
Whatever remains is at most a mass or gradient piece.
If that piece is nonzero, the linearized equation for $\Pi_i$ is algebraic and enforces $\Pi_i=0$.
That is why solving the Branch~II vector equations at linear order can appear to ``remove'' the St\"{u}ckelberg vector.
If the remaining quadratic term also vanishes, the action in this direction is empty and the situation is even more singular.
Neither case is a proof that the mode is absent from the full nonlinear theory.

The usual argument that $K_V=0$ implies infinite strong coupling comes from canonically normalizing the field.
Away from Branch~II one may define
\bea
    v_i
    \;=\;
    \frac{M_{\mathrm{Pl}}\, a k}{\sqrt{2}}\,\sqrt{K_V}\,\Pi_i \,,
\label{eq:canonical_vector}
\eea
so that the kinetic term in Eq.~(\ref{eq:S_V_kin}) takes a canonical form for $v_i$.
As $J\rightarrow 0$, one has $K_V\rightarrow 0$ and this field redefinition becomes singular.
Generic cubic (or higher) vertices that are regular in $\Pi_i$ then become interactions of $v_i$ enhanced by inverse powers of $K_V$.
This is how Ref.~\cite{Gumrukcuoglu:2017ioy} identifies infinite strong coupling at linear order.
Since minimal scalar or Abelian vector matter does not change $K_V$, it does not remove the singularity.

A vanishing quadratic kinetic term is not the same as showing that the field has been eliminated from the theory.
The calculation above is strictly linear, on an exact FLRW background.
At the next order, a fluctuation of the background, for example a perturbation of $X$ or of the metric, could multiply $\Pi_i'\Pi_i'$ and give $\Pi_i$ a kinetic term as soon as one leaves the exact Branch~II solution.
If that happens, the mode is still present in the theory; its kinetic term merely vanishes on this particular background, which is the usual strong-coupling situation.
Showing that the mode is truly gone would require a calculation beyond linear perturbation theory, which we have not done.
What we have shown is that neither a homogeneous canonical scalar nor a spectator Maxwell or Proca field can restore a nonzero $K_V$ on Branch~II.

The same remark applies to the scalar sector.
In Ref.~\cite{Kulchoakrungsun:2022zkk} the Branch~II scalar analysis produced a locking relation among the scalar perturbations and removed the longitudinal St\"{u}ckelberg field from the reduced linear equations.
Together with the two missing vector kinetic terms, this is suggestive, but a linear mode count cannot tell us whether Branch~II is strongly coupled or is a true constraint branch.
That question remains open.


\section{Summary and discussion}
\label{section:Discussion}

We have studied transverse vector perturbations in ghost-free quasidilaton massive gravity without a quasidilaton kinetic term, in vacuum and with two types of minimal matter.
The question was whether that matter can restore a quadratic kinetic term for the gravitational helicity-1 modes on the self-accelerating branch $J=0$.
 
In vacuum we recovered the result of Ref.~\cite{Gumrukcuoglu:2017ioy}.
After the auxiliary shift $B_i$ is integrated out, the kinetic coefficient of $\Pi_i$ is
\bea
    K_V
    & = &
    \left[
    1+\frac{k^2(r+1)}{2 a^2 J m^2 X}
    \right]^{-1} \, .
\eea
On Branch~I, $K_V$ can be nonzero and the ultraviolet no-ghost condition reduces to $J>0$.
On Branch~II, $K_V=0$ and both gravitational vector polarizations lose their quadratic kinetic term, so they are infinitely strongly coupled at this order.
 
A homogeneous canonical scalar has no transverse perturbation, so it cannot enter the vector action directly.
It does appear in the unsimplified $B_i$ constraint, through $V(\bar\chi)$ and $\bar\chi'$.
Those terms cancel after the Friedmann equation is used, and both $B_i$ and $K_V$ reduce to their vacuum expressions.
Thus a scalar can change the background evolution of $a$, $r$, and $X$, but not the map from those quantities to $K_V$.
 
A single Abelian vector is a better test, because it does have a transverse mode $A_i^T$.
Isotropy forces $\bar A_{\mu}=0$ for one such field, so the vector is a spectator and does not source the FLRW background.
At quadratic order there is then no mixing between $A_i^T$ and $(B_i,\Pi_i)$: every such vertex is at least cubic.
The gravitational calculation is identical to vacuum, and the kinetic matrix is diagonal.
A Proca mass, on the same background, gives a mass to $A_i^T$ but does not generate $B_i$--$A_i^T$ or $\Pi_i$--$A_i^T$ mixing, and does not change $K_V$.
On Branch~II the photon (or massive Proca vector) can propagate, while the gravitational vector kinetic term still vanishes.
 
The conclusion is the same in all three cases.
Minimal scalar, Maxwell, or Proca matter does not restore a nonzero $K_V$ when $J=0$.
At quadratic order the infinite strong-coupling problem of the gravitational vector modes therefore remains.
 
As discussed in Sec.~\ref{section:branchII}, this is a statement about the linearized theory on exact FLRW.
It does not prove that the gravitational vector modes have been eliminated from the nonlinear theory, nor does it replace a calculation beyond quadratic order.
It does show that simply adding ordinary, minimally coupled matter is not enough to make Branch~II perturbatively healthy in the vector sector.
 
Several extensions would be needed to go further.
A cubic (or fully nonlinear) analysis could test whether a kinetic term for $\Pi_i$ reappears once the background is allowed to fluctuate.
A nonzero isotropic vector background can produce quadratic mixing, but it requires several vector fields or a non-Abelian configuration.
Nonminimal couplings of matter to the fiducial metric or to the quasidilaton might change the $B_i$ constraint.
We have also not treated tensors, imperfect fluids with vector anisotropic stress, or a new scan of Branch~I parameters.
 
Taken together with Ref.~\cite{Kulchoakrungsun:2022zkk}, the linear status of the two branches is then as follows.
Both branches still exist in the presence of matter.
On Branch~I the gravitational vector kinetic term can be healthy if $J>0$, and the scalar sector can satisfy Higuchi-type bounds for a restricted, roughly pressureless potential, with scale-independent subhorizon growth.
On Branch~II the scalars lock at linear order and the gravitational vector kinetic term vanishes, with or without the minimal matter we considered.
For cosmological applications that need a weakly coupled linear perturbation theory, Branch~I remains the available option. The nonlinear fate of Branch~II is left open.


\acknowledgments

This research has received funding support from the NSRF via the Program Management Unit for Human Resources \& Institutional Development, Research and Innovation [grant number B13F670063]. DS has received funding support from the Fundamental Fund of Khon Kaen University. Generative AI tools (Grok, xAI; ChatGPT, OpenAI) were used to assist with drafting and revising the manuscript text. The authors directed this assistance, verified all scientific content and equations against their calculations, and take full responsibility for the paper.


\bibliography{eqmg}

\end{document}